\documentclass[11pt]{JHEP3}

\usepackage{latexsym, amssymb, amsmath, epsfig}

\newcommand{\sn}{$S_{n}$}
\newcommand{\snm}{$S_{n-1}$}
\newcommand{\sip}{$S_{n+1}$}

\newcommand{\spf}[2]{\Omega\left(#1,#2\right)}
\newcommand{\opf}[1]{\hat{\Omega}\left(#1,\cdot\right)}

\def\bea{\begin{eqnarray}}
\def\eea{\end{eqnarray}}
\def\be{\begin{equation}}
\def\ee{\end{equation}}
\def\beq{\begin{equation}}
\def\eeq{\end{equation}}

\def\a{\alpha}
\def\b{\beta}
\def\d{\delta}
\def\D{\Delta}
\def\e{\epsilon}
\def\g{\gamma}
\def\G{\Gamma}

\def\O{\Omega}
\def\o{\omega}
\def\s{\sigma}
\def\S{\Sigma}
\def\r{\rho}

\def\Oh{\hat{\Omega}}

\def\la{\langle}
\def\ra{\rangle}

\def\up{p^L}
\def\dop{p_L}
\def\pu{p^R}
\def\pd{p_R}
\def\lp{p_L^{\phantom{L}L}}
\def\pr{p_R^{\phantom{R}R}}

\title{Quantum field theory on a growing lattice}
\author{Brendan Z. Foster and Ted Jacobson\\
Institut d'Astrophysique de Paris,
98 bis Bvd. Arago, 75014 Paris\\
Email: \email{bzf@umd.edu, jacobson@umd.edu}}

\abstract{
We construct the classical and canonically quantized
theories of a massless scalar field on a background
lattice in which the number of points---and hence the
number of modes---may grow in time.
To obtain a well-defined theory certain restrictions
must be imposed on the lattice.
Growth-induced particle creation is studied
in a two-dimensional example.
The results suggest that local mode birth of this sort
injects too much energy into the
vacuum to be a viable model of cosmological mode birth.}


\preprint{}

\begin{document}

\section{Introduction}
\label{intro}

If, as suspected by many,
there is a short distance cutoff on physical
degrees of freedom, then as the universe
expands the number of degrees of freedom must
be growing. Such a scenario might neatly account
for the otherwise puzzlingly low initial entropy
of the universe, and at long wavelengths might
be indistinguishable from the usual continuum
physics in an expanding background spacetime.
But to explore this avenue is difficult,
since it raises the problem of formulating dynamics
in a setting where the number of degrees of freedom is
not the same at each time~\cite{Jacobson:1999zk}.

In the framework of effective field theory (EFT),
it is straightforward to describe mode birth.
Consider by analogy a growing droplet of liquid.
The collective vibrational modes define an EFT
with a cutoff at the atomic scale. As new
atoms condense onto the droplet, new modes are added;
but at the growing scale, the EFT does not apply.
One can work in the Fourier
transform domain, not with a distance cutoff but with a
wavevector cutoff. Then EFT modes can be
inserted by hand in the adiabatic ground state, but
in this approach one learns
nothing about the microscopic physics associated with
growth of the microstructure of spacetime.

Perhaps a quantum gravity theory can be formulated
in which states with different numbers of degrees of freedom
are all contained in the same Hilbert space, so that
in the pre-geometric sense the number of degrees of
freedom is actually constant. Then presumably mode
birth would be just a feature of the dynamics.
This is a deep problem.
Here we try to gain some insight by considering what
should be a much simpler problem:
quantum field theory (QFT) on a background growing,
or more generally irregular lattice.

An appealing idea for a fundamental discrete, irregular
structure is that of a
{\it causal set}~\cite{Bombelli:aa,Sorkin:2003bx}.
It is an outstanding problem whether and how one
might formulate quantum field theory on a causal set.
(Some ideas on classical field theory have been
worked out however~\cite{Daughton:1993rh}.)
This question forms a second motivation for
the present work, although the causal sets we
consider have two properties not shared
by a generic causal set: finite linking and a
`layered' structure.

We thus wish to consider a simple free scalar
field theory model on an irregular lattice structure. The
challenge is to formulate a theory that allows for
the increase of degrees of freedom, while possessing a
good quantum mechanical interpretation and agreeing
at long wavelengths with standard quantum field theory
in an expanding background spacetime.

\subsection{Overview of the approach and results}

It is by no means clear how to modify quantum field theory
to allow for irregular lattices, in particular growing ones.
One could imagine just writing down an action, and
a corresponding path integral.
This has been done on random lattices in
Euclidean signature, and the path integral can
be given there a statistical probability
interpretation. But in the Lorentzian context,
without the usual
underlying Hamiltonian structure there is no guarantee that
the complex amplitudes so defined would have any consistent
probability interpretation at all.  For example the transition amplitudes
would in no sense be unitary.
One could try to use the decoherence functional approach,
rather than demanding strict unitarity.
In fact just such an approach was explored
in~\cite{Blute:2001wp} in a very general
setting of evolution on causal networks. But there is no guarantee
that useful sets of decohering histories
could be found in such a general setting.
Abstract formulations of quantum dynamics on causal sets
have also been developed\cite{Blute:2001zd,Hawkins:2003vc},
but without direct
relation to the dynamics of any particular field theory.
For the present work the aim is simply to recover the
quantum field theory of a linear field in the large,
so we prefer to
stay closer to a local Hamiltonian formulation,
adopting the algebraic approach to QFT\cite{Haagbook, Waldbook}.
It may be that for the class of lattices considered here, the path integral
approach {\it does} make sense, but we have not investigated that question.

The algebraic approach has for ingredients a $C^*$-algebra and
a state. For example, the algebra might be
generated by the abstract quantum field
``operator" satisfying a field equation and
canonical commutation relations,
and the state is a positive linear functional on the algebra
that assigns expectation values to observables.
The conservation of a symplectic structure under
evolution from one time slice to the next
plays a critical role in the
way both of these structures are usually defined in practice.
The equation
of motion must be compatible with the commutation relations,
a property whose validity derives ultimately from the conserved symplectic
form in phase space. And candidate states are defined as
elements of the Fock space built over a one particle Hilbert
space whose inner product  also derives from the
conserved symplectic form.

While we can loosen the form of the equation of motion,
it is difficult to know where to begin if we do not maintain
the existence of a conserved symplectic form. Thus, at least
for now, we restrict to irregular lattices that nevertheless possess
a time layering so that the dynamical connection between points
proceeds systematically from one time slice to the next,
allowing a conserved symplectic form to be defined.

Even when there is a conserved symplectic form, the
equation of motion we write will generally not admit an
initial value formulation, as evolution from initial data may
not be unique or may not exist in general. This is inevitable
when the number of field points and/or their connectivity
is not conserved from one
time to the next, since there can be either too few or too many
equations to determine the subsequent values. One of the
interesting questions is just how far such behavior can be tolerated.

In order to
implement the commutation relation and define the
Hilbert space in a straightforward way we
restrict the lattice so that a linear phase space
consisting of solutions to the field equation can be
defined. This requires that backwards evolution,
if it exists, is unique.
To include all observables of interest,
it is necessary to allow for `birth solutions' that come into
existence at a finite time, i.e. that
cannot be evolved backwards past a certain time.
With all of this structure we find it possible
to realize the algebraically-defined
quantum field theory in a way that preserves the equations of motion,
has canonical commutation relations, and allows for mode birth.
The ambiguity of future evolution is incorporated naturally
into this structure and does not present any fundamental problem.
We also construct a Fock representation of the algebra, with the help
of a particular rule for resolving the evolution ambiguity.
The Fock representation depends on this rule, but all such
representations are equivalent, and the theory is fully defined
by the algebra without the choice of any such rule.
The state of the newborn modes is not specified by the
earlier data so is extra information not available
at an earlier time.

We study here particle creation in a 1+1 dimensional
example, assuming the usual vacuum on an early time
regular lattice and examining the spectrum of created
particles in a late time
regular lattice after points have been added.
Considering all states subject to the restriction
that there are no incoming particles, we arrive at
the conclusion that,
roughly speaking, one Planck unit of energy is injected per
Planck volume (really length in the 2d model) created.
This calculation serves to illustrate that the theory has
been defined well enough to carry out such a computation.
But it also suggests that we have not achieved a viable
theory: it would
be incompatible with the observed universe to have so
much energy injected. We return to this issue
in the discussion section.

Other approaches to the problem of formulating mode birth in an
expanding universe have been explored recently.
Kempf~\cite{Kempf:2000ac} has constructed a theory of fields with
finite density of states, derived by representing a deformed
uncertainty relation between position and momentum which
implements a minimum spatial uncertainty. A quadratic scalar field
Lagrangian written with this deformed momentum operator produces a
field equation whose mode solutions are born at a definite time.
The modes cannot be evolved backward before the birth time,
because the differential equation they satisfy is singular at that
time. This is closely analogous to what occurs in our discrete
model. Unlike our case however, forward evolution is not
ambiguous.

In a very different approach, Sasakura~\cite{Sasakura:2004vm}
formulates discrete field theory of scalars, spinors, and gauge
fields, on a growing fuzzy 2-sphere. At each discrete time $N$ the
scalar field is an hermitian operator on the representation space
of the spin-$N/2$ representation of $su(2)$, and the fields
$\phi_{N+1}$, $\phi_{N}$, and $\phi_{N-1}$ are related by a
discrete analog of the wave equation. The classical scalar field
equation in an expanding universe with proper time $N$ and scale
factor $\sqrt{N+1}$ is recovered in a large $N$, low spin limit.
The initial value problem is not discussed, but it seems that the
field equation does not determine $\phi_{N+1}$ from $\phi_{N}$ and
$\phi_{N-1}$. We would guess that there are also birth solutions.
The quantum interpretation of this theory is not yet entirely
clear (at least to us). Both of these approaches share the feature
that a short distance cutoff exists without sacrificing spatial
symmetry, unlike in our rather more crude approach.

The rest of this paper is organized as follows. Section 2
describes the lattice structure adopted, and section 3 formulates
the action and field equation, as well as the phase space
description. The ingredients then allow straightforward quantization,
as described in section 4. In section 5 we work out the example of
quantization on a regular `hyperdiamond' lattice, and section 6
deals with the example of a two-dimensional lattice that grows
by the addition of two points.
In particular we construct a Fock space representation and
examine the resulting particle creation.
A discussion of the results, some perspectives, and
ideas for future work are given in section 7. An appendix
contains formulae for the Bogoliubov coefficients in the
two-dimensional example. We adopt the metric signature
$({+}{-}{-}{-})$
and use units with the speed of light $c=1$.

\section{Lattice structure}

We use a
type of lattice that is a special kind of
causal set~\cite{Bombelli:aa,Sorkin:2003bx}.
This choice allows the discrete structure
to carry the metric information in its topology,
with no length assignments.
To make the field theory action
finite, we restrict to the case where
each point $p$ has a finite number of causal neighbors,
i.e. points  $q$ such that there is no point between
$p$ and $q$ in the sense of the causal order.
We denote by $IF(p)$ the set of future causal neighbors,
the {\it immediate future}. Similarly $IP(p)$ refers to the
{\it immediate past}. These sets are thought of
as discrete analogs of a cross section of
the future and past light cone
in the tangent space at a point in the continuum.
A pair of causal neighbors is called a {\it link},
so the structure could be called a {\it finitely linked
causal set}.

We choose to preserve symmetry among the future
links at a point,
in the sense that our field theory action
is unchanged when they are permuted.
In the continuum limit
this amounts to treating the future
links as spacetime vectors of
identical length and with identical pairwise
inner products. Unlike the standard interpretation of
the ordering relation in causal sets, the links
will correspond to spacelike rather than null vectors
(except in two dimensions),
although the convex hull spanned by these vectors
will have null faces.

The action and equation of motion will be
well defined on the minimal structure of a finitely linked causal set.
But to make progress on the quantum field theory,
as discussed in the introduction,
we shall require that the lattice be layered by time
slices:
the lattice will consist of an indexed
sequence $\{S_n\}$ of sets of points called
{\it slices}, such that every point
lies in exactly one slice $S_n$, and links connect points in
slice \sn\hspace{.3pt} only to points in \snm\hspace{.3pt} or
\sip.\footnote{In the causal set literature a ``slice" refers to
a maximal set of unrelated points. Our slices here consist of
unrelated points, but they need not be maximal. For example
the points $e$ and $f$ in Fig.~\ref{birth} are unrelated to points
on the previous slices, so the latter are not maximal.}

We will use the letters
$i,j$ to index elements $p^i\in IF(p)$ and $p_i\in IP(p)$ with
no implication that the number of elements in any of these
sets is equal. Multiple indices are also allowed,
thus for example
\beq
p_i{}^j:=(p_i)^j\in IF(IP(p)).
\eeq
Note that the index order matters: $p^j{}_i\in IP(IF(p))$
is generally not equal to $p_i{}^j$. One calls $p^i$ a {\it child}
and $p_i$ a {\it parent} of $p$. Then $p_i{}^j$ is a
{\it sibling} while $p^j{}_i$ is a {\it mate}. All manner
of polygamy and incest is permitted by the layered
structure, provided it is not cross-generational.

To carry out the quantization in close analogy with
linear quantum fields in the continuum, we will make one further
restriction on the lattice structure: backward-in-time evolution
using the field equation we adopt, if it exists, is unique.

\subsection{Hyperdiamond lattice}
\label{hd}

The regular prototype of the lattices we will
consider is a $d$-dimensional hypercubical space-time lattice
tilted on its vertex so that a
preferred time-like direction runs from one corner of the cube to
the complete opposite.
We call this the {\it hyperdiamond lattice} after the
terminology of Ref.~\cite{Finkelstein:1996wu,Smith:1995kd}.
A step along one lattice link runs along
one of the cube's edges, and
represents a movement in the preferred time-like direction and a
movement in the orthogonal space
to one of the $d$ vertices of a
$(d-1)$-dimensional tetrahedron centered on the original point.
The lattice thus possesses a tetrahedral rotation symmetry.

We choose the step-speed so that the future continuum light cone
at a point lies inside and tangent to the polyhedron
formed by the set of future links. With this choice
the hyperplane faces of the polyhedral cone are null
hyperplanes and the links are spacelike.
This
determines the step speed as follows.

The $d$ link vectors $n_i^a$ are given by
\beq
n_i^a = t^a +\a s_i^a
\label{nia}
\eeq
where $t^a$ is a unit timelike vector along the hypercube diagonal
and, for each $i$, $s_i^a$  is a spatial unit vector orthogonal to
$t^a$ pointing to one of the vertices of a $(d-1)$-dimensional
tetrahedron. By tetrahedral symmetry, the sum $\sum_i s_i^a$ must
vanish. If the cube faces are to be null, then their normal vector
must lie within them and must be a null vector. Again by
tetrahedral symmetry this normal for the face omitting $n_i^a$
must be the sum $\sum_{j\ne i}n_j^a$ of the $d-1$ link vectors
defining the face. Since the $d$ spatial vectors sum to zero this
is equal to $(d-1)t^a -\a s_i^a$, which is null provided $\a=d-1$.
Hence the link step speed is $d-1$, which is the speed of light
only in two spacetime dimensions.

Squaring the relation $\sum_i s_i^a=0$ reveals
the spacetime inner products of the spatial vectors,
$s_i^as_{ja}=1/(d-1)$ for $i\ne j$. With our signature,
the spatial inner products have the opposite sign, so the
cosine of the angle between any two is $-1/(d-1)$.
The inverse spacetime metric can be expressed in terms of the
link vectors as
\beq
g^{ab}=\frac1{d(d-1)}\sum_{i\ne j} n_i^a n_j^b.
\label{metric}
\eeq

An equivalent description
of the hyperdiamond lattice is obtained
beginning with a system
of $d$ null coordinates $v^i$, i.e. coordinates whose level
sets are null hyperplanes, such that the metric is
invariant under permutations of these coordinates.
The corresponding basis one-forms $dv^i$ are null, and
the dual basis vectors $(\partial/\partial v^i)^a$
are equal to the $n_i^a$ defined in (\ref{nia}),
when the $v^i$ are scaled so that $t=\sum_i v^i$ is
a proper time coordinate.
Discretizing the $v^i$ then yields the lattice.

The $d$-dimensional
hyperdiamond lattice can be
compactified to have spatial sections that
are $(d-1)$-tori. Consider for illustration the case
$d=4$. Then the immediate future of any point is a tetrahedron.
Translation along any edge of the tetrahedron is clearly
a symmetry of the hyperdiamond lattice, since every point is
equivalent. We can thus identify the points related by
any integer number of such translations along three
edges at one vertex to obtain
a 3-torus. One can also compactify to ``twisted tori", or even
Klein bottles.

\section{Lattice dynamics}

A continuum massless, minimally coupled
scalar field in $d$ spacetime dimensions
has the action
\beq
S=\tfrac12\int d^dx \sqrt{-g}g^{ab}\partial_a\phi\partial_b\phi.
\label{contact}
\eeq
We first discretize this on a hyperdiamond lattice,
then use the result to motivate an action on an irregular
lattice. After discussing the resulting equations of motion, we
introduce  a phase space and symplectic form to be used in
the quantization.

\subsubsection{Hyperdiamond lattice action}
Given a hyperdiamond lattice with edge links
$\e n_i^a$ scaled by a small length $\e$, we insert the
metric decomposition (\ref{metric}) in the action
(\ref{contact})
and replace the partial derivatives by finite differences
\beq
\e n_i^a\partial_a\phi\rightarrow
\phi(p+\e n_i)-\phi(p).
\eeq
This yields the discrete action
\beq
S_{\rm reg}=
V^{(d)} \e^{d-2}\tfrac12
  \sum_{\substack{p\\ i\ne j}}
    \frac1{d(d-1)}\, \Big(\phi(p^i)-\phi(p)\Big)\Big(\phi(p^j)-\phi(p)\Big),
\label{regact}
\eeq
where $V^{(d)}\e^d$ is the spacetime volume of one
hypercubical cell of the lattice, and $p^i$ denotes
the point future-linked to $p$ in the direction $n_i^a$.

One could instead use the past
links rather than the future links, or even a combination
of the two, in the discrete action. We use the future
links for no good reason.

\subsubsection{Irregular lattice action}

Using the regular case (\ref{regact}) as a guide,
we posit an action for a scalar field on a
general causal lattice. We cannot copy that
exactly at each point $p$, since the number of
points $d_p$ in the immediate future
$IF(p)$ varies from point to point, hence
different powers of the dimensionful quantity
$\epsilon$ appear. Our definition of
the discrete theory simply ignores
the powers of $\epsilon$ and the volume factor
$V^{(d)}$  that goes along with them. We
preserve the denominator $d_p(d_p-1)$, which
corresponds to the number of terms in the sum,
so that the action at $p$ is always the average of the
products of the differences along the future links at $p$.
With these choices, the irregular lattice action we adopt is
\beq
S_{\rm irreg}=
\tfrac12
  \sum_{\substack{p\\ i\ne j}}
        \frac{1}{d_p(d_p-1)}\,
    \Big(\phi(p^i)-\phi(p)\Big)\Big(\phi(p^j)-\phi(p)\Big).
\label{irregact}
\eeq

\subsection{Discrete field equations}

Varying the field in the irregular lattice action (\ref{irregact})
one obtains the discrete field equation
\be
\Box\phi = 0,
\label{waveeqn}
\ee
where the discrete D'Alembertian operator $\Box$ is defined by
\be
\label{BOXD1}
    \Box\phi(p)  = -\phi(p)
   +\sum_{i}\frac{1}{d_p}\,\phi(p^i)
    +\sum_{i}\frac{1}{d_{p_i}}\,\phi(p_i)
     - \sum_{\substack{i,j\\ p_i{}^j\neq p}}
            \frac{1}{d_{p_i}(d_{p_i}-1)}\,\phi(p_i{}^j).
             \ee
In the regular case, the D'Alembertian is the
average over the children, plus the average over the parents,
minus the average over the siblings, minus the
value of the field at $p$. It does not have this interpretation
in the general case, because the weights of the
parent and sibling terms are different.
It can be rewritten to look more like the continuum
D'Alembertian $\partial_t^2-\nabla^2$:
\be
\label{BOXD2}
   \sum_{i}\frac{1}{d_p}\,\Big(\phi(p^i)-\phi(p)\Big)-
        \sum_{i}\frac{1}{d_{p_i}}\,\Big(\phi(p)-\phi(p_i)\Big)
        -\sum_{i,j}
        \frac{1}{d_{p_i}(d_{p_i}-1)}\,\Big(\phi(p_i{}^j)-\phi(p)\Big).
\ee
In the regular case the first two sums are the
average future and past time-differences, so their
difference gives a discrete version of the second
time derivative. The third sum is the average of
the spatial neighbors minus the value at $p$,
which is a discrete version of the spatial Laplacian.

\subsection{Initial value problem}
\label{ivp}

The field equation (\ref{waveeqn},\ref{BOXD1}) involves values
of the field on three consecutive time slices.
The equation that arises from varying $\phi(p)$
on the middle slice determines the average value on
$IF(p)$ given the values on the first two slices.
If the number of points on the middle and third slice
is $N_2$ and $N_3$ respectively,
then one has $N_2$ linear equations in the
$N_3$ unknown values on the third slice.
Supposing $N_2$ and $N_3$ are finite,
the future solution is not uniquely determined if $N_3>N_2$,
and there is generically no future solution if $N_3<N_2$.
Existence of a future solution when $N_3\ge N_2$ depends
upon the connectivity of the lattice.

An example we shall make use of
is a two-dimensional regular diamond
lattice, where each point has two parents and two
children, and periodic boundary conditions are imposed
so that at each time there is a total of $N$
points arranged on a circle.
The field equation determines
the sum of the values on each adjacent pair of points,
so if $N$ is even (or infinite) the evolution
is determined only up to the addition of a
spatially-alternating function with arbitrary
time-dependence. If $N$ is
odd the evolution is uniquely determined since
the periodicity is inconsistent with the alternation.
It is possible
to regard the alternating functions in the even or infinite
cases as ``pure gauge",
and to formulate a corresponding dynamics of the
gauge-invariant degrees of freedom,
but we shall avoid this extra complication by requiring
$N$ to be odd in the two-dimensional example we
study.

Note that even if the number of points on each slice is
the same, the equations could be over-determined so that
no future solution would exist. For example this could
happen if the future of some
set of points on one slice contains fewer points
on the next slice. But even if these numbers
match there could still be trouble. Consider the two-dimensional
example just discussed in the case that $N$ is even.
The equations take the form $\phi(i)+\phi(i+1)=H_{i}$, with
$i=1,\dots,N$ and $\phi(N+1)\equiv\phi(1)$.
The sum of the left hand
sides of the odd-$i$ equations
is equal to the sum of the even-$i$ ones.
(For example, with $N=4$ one has $[\phi(1)+\phi(2)]+[\phi(3)+\phi(4)]=
[\phi(2)+\phi(3)]+[\phi(4)+\phi(1)]$.) This implies that there can be
a solution only if the condition $\sum_{\rm odd}H_i=\sum_{\rm even}H_i$ holds.
It happens that this condition is automatic if the lattice is
a regular diamond lattice, but it could fail to hold
on a lattice with altered connectivity prior to the slice in question.

\subsection{Phase space of half-solutions}
\label{Gamma}

Since initial data cannot in general be evolved
indefinitely or uniquely, it is not convenient to define
a phase space in terms of initial data. Instead one can try
to use the covariant approach that identifies phase space as the
space of solutions.\footnote{See~\cite{Waldbook} for an introduction
to the covariant phase space formalism and its application
to quantization of linear field theory.}
This takes care of non-uniqueness of evolution,
as two distinct evolutions simply correspond to different
points in phase space. Non-existence of evolution is
another matter.

One might try to associate to each region
a different phase space, consisting of
the solutions that are defined everywhere in that
region. This idea may be compatible with that
of a net of local algebras on which the algebraic formulation
of QFT is based\cite{Haagbook}. Another approach might be
to introduce just one phase space, without a linear structure.
But, for this initial investigation, we
choose to exploit the linear structure of a single
phase space in strict analogy with standard linear
quantum field theory.
Thus we seek a definition of phase space as a single vector space.
In particular, we require an operation of addition of two solutions
that defines an abelian group, i.e. that
is commutative, associative, and possesses an identity and
an inverse.  This requirement leads to further demands
on the partial solutions and lattice structure.

Let $\phi(s..t)$ denote a solution that exists between the
discrete times $s$ and $t$ and is undefined outside that interval.
The additive inverse of $\phi(s..t)$ must surely be
$-\phi(s..t)$, but these two functions are defined only on the
interval $[s,t]$, whereas the identity must be the zero
solution which exists for all time. We thus learn that
the operation of addition must include a maximal extrapolation
(using the field equation) beyond the original range of definition of the
sum. However, since evolution is generally not unique, extrapolation
is not well-defined.
To cope with this we restrict to lattices for which
{\it backward} evolution, if it exists, is unique,
and we define the phase space to consist of
solutions that are defined for all time,
together with solutions that are defined on
a semi-infinite interval\footnote{The theory could also
be set up using solutions defined on intervals of the form
$[t,t_0]$, where $t_0$ is a fixed finite time.}
$[t,\infty)$.
These latter we call {\it half-solutions}
with {\it birth time} $t$.
On this space,
the operation of  ``sum on the common domain, then maximally
extend to the past" defines an abelian group.
To see that this is associative consider
$(a+b)+c$ and $a+(b+c)$.
In the region where $a$, $b$, and $c$ are all individually
defined these sums are both equal to the standard sum $a+b+c$.
Globally, they are both the maximal past extension of this sum,
hence they are equal.

As noted in section (\ref{ivp}),
there is no guarantee that all---or even any---initial data can be
extended indefinitely into the future. Despite this we can
define the phase space using the half-solutions. If there are none,
then the theory is empty.

\subsection{Symplectic form}

In a general Hamiltonian
context, a symplectic form $\O$ on phase space is
a non-degenerate closed two-form~\cite{Waldbook}.
For a linear phase space $\G$ the tangent space at a
point is naturally identified with the phase space
itself, so $\O$ can be treated
as a map from $\G\times\G$
to the real numbers. In the present application,
$\O$ maps pairs of half-solutions to numbers.

To motivate the discrete version, recall first how it works
in the continuum.
For a real Klein--Gordon field in
$D$-dimensional space-time, the symplectic
product between two solutions $\psi$ and $\phi$
is defined by
\be
\label{SIMP}
    \spf{\psi}{\phi} = \int_S\bigl(\psi\,
    \partial_a\phi - \phi\partial_a\psi
   \bigr)\,d\Sigma^a,
\ee
where $S$ is a Cauchy surface.
This product is well-defined on the
covariant phase space since
it takes the same value when evaluated on any surface
$S$ provided the field equations are satisfied.
A simple way to verify this fact is to integrate the expression
$\psi\Box\phi-\phi\Box\psi$, which vanishes by virtue of
the field equation,  over a space-time region between
two Cauchy surfaces. This integrand is the total divergence
$\nabla^a(\psi\,\partial_a\phi - \phi\partial_a\psi)$,
hence the volume integral is equal to the difference
of (\ref{SIMP}) between the final and initial surfaces (assuming
no contribution from a spatial boundary). Since the volume
integral vanishes, the two spatial integrals must be equal.

The same construction works in the discrete case,
with $\Box$ replaced by the discrete version (\ref{BOXD1}).
The sum of $\phi\Box\psi - \psi\Box\phi$
over a sequence of slices reduces by cancellations to  a pair of
boundary contributions on the initial and final slices (assuming
no contribution from a spatial boundary). This
shows that the discrete symplectic form
\beq
\O(\psi,\phi)=\sum_{p\in S;\, i}\frac{1}{d_p}
\bigl(\psi(p)\phi(p^i)-\phi(p)\psi(p^i)\bigr)
\label{Omega}
\eeq
takes the same value when computed on any slice $S$
in the domain of definition of two solutions
$\psi$ and $\phi$ to the discrete field equation.
It is thus well-defined on phase space.
On a spatially infinite lattice some kind of falloff
condition on the fields would be required to
ensure that the infinite sum defining the
symplectic form converges.

\subsubsection{Degeneracy}

The symplectic form for the continuum Klein--Gordon
field is non-degenerate, i.e. $\O(\psi,\cdot)=0$ only if
 $\psi=0$. This is not necessarily true on the lattice.
For example, suppose $A$ is any spatially alternating
configuration on a two-dimensional periodic even diamond lattice.
Then,  as discussed in section \ref{ivp}, $A$ satisfies
the field equation. It is thus an element
of the phase space, so $\O(A,\cdot)$ is well
defined, and is in fact equal to zero. Whenever a symplectic
form has a vector field $Y$ for which $\O(Y,\cdot)=0$,
the Hamiltonian vector field $X_H$ is determined by
Hamilton's equations $dH=\O(\cdot, X_H)$ only up to
addition of an arbitrary multiple of $Y$. In the present
case, this corresponds to the statement that to any
solution $\phi$ one can add an arbitrary spatially-alternating
configuration to obtain a new solution with
the same initial data. This indeterminacy is a gauge
freedom, and the physical phase space is the quotient
of the solution space modulo the gauge freedom.

\subsection{Poisson brackets}

The symplectic form, if not degenerate,
provides the definition of Poisson brackets
between any two
phase space observables, i.e. functions on phase space.
Given two such functions $f$ and $g$,
the Poisson bracket is a third function defined by
\beq
\{f,g\}=\O^{ab}\partial_a f\, \partial_b g.
\label{PB}
\eeq
Here we use (abstract) index notation,
and $\O^{ab}$ is the inverse of the symplectic
form $\O_{ab}$, defined by $\O^{ab}\O_{bc}=\d^a_c$.

In the linear case, to each phase
space point $\phi$ is associated the
linear observable whose value at $\psi$ is
$\O(\phi,\psi)=\O_{ma}\phi^m\psi^a$. This function
has the gradient
$\partial_a\O(\phi,\cdot)=\O_{ma}\phi^m$; therefore
\beq
\{\O(\psi,\cdot),\O(\phi,\cdot)\}=-\O(\psi,\phi).
\label{PB2}
\eeq
It is this relation
that we quantize. If $\O$ is degenerate,
an algebra of gauge-invariant observables
can be defined directly by the
relation (\ref{PB2}),
which does not involve the inverse of $\O$.

\subsubsection{Local fields and locality}

Though not central to our study, we shall make a few
remarks here about local observables and their Poisson brackets.
This section can be omitted without missing anything
essential to the central focus of our paper.

First, note that evaluation of solutions at a lattice point
$x$ defines a linear map $\phi\rightarrow\phi(x)$ only on those
half-solutions whose birth time lies before $x$.
Since this map is not defined on the entire phase
space it is not strictly speaking an observable.
The Poisson bracket $\{\phi(x),\phi(y)\}$
is thus not globally defined on the phase space, but
it should be meaningful locally in phase space.

To discuss the locality in space-time of the
Poisson brackets, let us restrict to the
case of a compact $d$-dimensional hyperdiamond lattice, chosen
so that evolution from initial data always
exists and is unique. Then the local field
$\phi(x)$ is an observable.
We call this a {\it deterministic lattice}.

In the continuum, the Poisson bracket of two local fields
is given by
\be
\label{UNBR}
    \{\phi(x),\phi(y)\}
    = G_R(x,y) - G_A(x,y),
\ee
where $G_R$ and $G_A$ are the retarded and advanced Green's
functions of the Klein--Gordon equation. If $x$ and $y$
are spacelike related this bracket vanishes. On the
lattice (in greater than two dimensions) the links are spacelike,
so we do not expect the bracket to vanish at all spacelike
separations. However, not even a lattice locality
property survives, as we now explain.

On a deterministic lattice
$\O$ is necessarily invertible,\footnote{If
$A$ is a solution such that
$\O(A,\cdot)=0$, then the
future and past averages of $A$ must vanish at every point.
The field equation (\ref{waveeqn},\ref{BOXD1}) then implies that $A=0$,
unless $d=2$. In the case $d=2$, uniqueness of evolution requires
an odd number of points on each slice, which together with
the vanishing of averages implies $A=0$.}
so there exists a solution $F_x$ such that
\beq
\phi(x)=\O(F_x,\phi).
\label{Fx}
\eeq
Evaluating on the slice $S_x$, one sees that
in order for Eq.~(\ref{Fx}) to hold for all $\phi$,
the solution $F_x$ must vanish everywhere on the slice $S_x$
containing $x$, have a mean value of $-1$ on
$IF(x)$, and have a mean value of zero on $IF(v)$ for
every other point $v$ in the slice $S_x$.

The bracket of local fields is thus
\beq
\{\phi(x),\phi(y)\}=-\O(F_x,F_y)=-F_y(x).
\label{xy}
\eeq
The second equality follows
immediately from the defining property
(\ref{Fx}) of $F_x$.
If $x$ and $y$ lie on the same slice then
$F_y(x)=0$, so the equal-time
bracket vanishes, as in the continuum.
If $y$ lies on the subsequent slice,
we need to use the field equation to
evolve $F_y$ back one step to the slice $S_x$
in order to evaluate $F_y(x)$.
The field equation (\ref{waveeqn},\ref{BOXD1})
on $S_y$ implies that the
average of $F_y$ on $IP(y)$ is $+1$ and the average on
$IP(w)$ vanishes for every other point $w$ in the slice $S_y$.
On a hyperdiamond lattice
this clearly forces $F_y$ to have support on
$S_x$ arbitrarily far from $y$, hence there will
be many such pairs $x$, $y$ for which $F_y(x)\ne0$.

This nonlocality of the bracket (\ref{xy})
survives quantization, so it
implies arbitrarily spacelike distant observables
fail to commute. It is not clear that this
is really a problem, however, since the {\it average}
of field values on each $IF(x)$ may have purely local brackets.
This average corresponds to the observable
$\O(\d_x,\cdot)$, where $\d_x$ is equal to $1$ at
$x$ and vanishes everywhere else
on the slice $S_x$ containing $x$ and the subsequent
slice. That is, we have
\beq
\frac1{d}\sum_i \phi(x^i)=\O(\d_x,\phi),
\label{deltax}
\eeq
so the bracket of two such local averages is
\beq
\Bigl\{\frac{1}{d}\sum_i \phi(x^i),\frac{1}{d}\sum_j \phi(y^j)\Bigr\}
=-\O(\d_x,\d_y)=-\frac{1}{d}\sum_i \d_y(x^i).
\label{xyavg}
\eeq
Unlike for $F_x$, the data defining $\d_x$ is
localized. Since the field equation (\ref{waveeqn},\ref{BOXD1}) is
local this means that the support of the
solution $\d_x$ {\it might} propagate ``causally"
away from $x$ to the future and the past. If so,
then the bracket of the averages
is non-zero only if $x$ and $y$ are causally related in the
sense determined by the local field equation. On an
odd two-dimensional lattice the propagation of $\d_x$ is
indeed causal. We do not know if that property holds
in higher dimensions.

\section{Quantization}
\label{Q}

To quantize one replaces the classical
observable $\O(\psi,\cdot)$ by an element
$\Oh(\psi,\cdot)$ of a $C^*$-algebra possessing
the canonical commutation relations (CCR),
\be
\label{ALG}
    \left[\opf{\phi},\opf{\psi}\right] = -i\hbar\spf{\phi}{\psi}\hat 1,
\ee
and such that $\Oh(\psi,\cdot)^*=\Oh(\psi,\cdot)$.
(Hereafter we set $\hbar=1$.) The quantum theory is then
completed with the choice of a state, i.e. a linear
functional $\o$ on this algebra satisfying the
positivity
condition $\o(a^*a)\ge0$, with $a$ an arbitrary algebra element,
and the normalization condition $\o(1)=1$.
This notion of state adapts
the concept of a density matrix to the most general setting.

Here we follow the simple conventional procedure of
constructing the class of states by representing the
algebra as operators on a Fock space.
The Fock space is
built over a ``one-particle" Hilbert space which
is defined using the {\it Klein--Gordon inner product} between
complex solutions to the classical field equation:
\beq
\la\psi,\phi\ra=i\O(\bar\psi,\phi)=i\sum_{p\in S;\, i}\frac{1}{d_p}
\bigl(\bar\psi(p)\phi(p^i)-\phi(p)\bar\psi(p^i)\bigr).
\label{KGip}
\eeq
The definition (\ref{Omega})
of $\O$ is extended here by complex linearity
to the space of complex solutions to the field equation.
Although this is not a positive-definite inner product
it can be used to construct
the Hilbert space as follows.

Choose a positive-norm subspace which is orthogonal
to its conjugate, and which together with its conjugate
spans the space of solutions.
Let $\{\xi_i,\bar\xi_i\}$ be an orthonormal basis adapted
to this decomposition, so that
\beq
\la \xi_i,\xi_j\ra=\d_{ij}=-\la \bar\xi_i,\bar\xi_j\ra,
\qquad \la \xi_i,\bar\xi_j\ra=0,
\eeq
and define the field operator $\hat\phi$ as
\beq
\hat\phi = \sum_i \bigl(\xi_i a_i + \bar\xi_i a^\dagger_i\bigr),
\label{fieldop}
\eeq
where
\beq
[a_i,a^\dagger_j]=\d_{ij}, \qquad [a_i,a_j]=0=[a^\dagger_i,a^\dagger_j].
\label{aadagger}
\eeq
Then the commutation relations (\ref{ALG}) are satisfied if
the quantized linear observables are represented
by
\beq
\bigl[\hat\O(\psi,\cdot)\bigr]_{\rm rep}=i\O(\psi,\hat\phi)=
i\sum_i \bigl(\O(\psi,\xi_i) a_i + \O(\psi,\bar\xi_i) a^\dagger_i\bigr).
\eeq
The `Fock vacuum' state $|0\rangle$ is defined by the
condition $a_i|0\ra=0$ for all $i$, and the
Fock space is the Hilbert space spanned by
all finite norm vectors obtained by acting
on $|0\rangle$ with products of the raising
operators $a_i^{\dag}$. The field operator
(\ref{fieldop}) is self-adjoint acting on
the Fock space, as befits a real field.
The state vectors of the theory are the elements
of Fock space, while the algebraic states are density
matrices constructed therefrom.

The construction in the last paragraph
depends upon the basis $\{\xi_i,\bar\xi_i\}$,
but on a spatially finite lattice
the representation of the algebra is independent
of this choice. This follows from
the Stone--Von-Neumann theorem,
which states that all representations
of the canonical commutation relations
for a finite dimensional phase space are
unitarily equivalent\cite{Waldbook}.

\section{QFT on a hyperdiamond lattice}
\label{qft}

In this section we work out the example of a
toroidally compactified hyperdiamond
lattice, beginning with the classical fields and
then quantizing. We
initially consider
an arbitrary dimension $d$.
We find that the discrete field equation
has exponentially growing modes
except in two dimensions.
In the next section we use these
results in the two-dimensional case to
determine the particle creation on a growing lattice.

\subsection{Hyperdiamond modes}
\label{sec:hdmodes}
We can use the harmonic modes
\beq
\eta_k = {\cal N}_k e^{-ik_i v^i},
\label{hdmodes}
\eeq
where $v^i$ are the null coordinates discussed
at the end of section \ref{hd}, and ${\cal N}_k$
is a normalization factor. We adopt units with
the lattice spacing $\epsilon=1$, so one future link on
the lattice corresponds to adding unity to one of the
$d$ null coordinates $v^i$. The toroidal boundary conditions
select a set of allowed wave vectors $k_i$.

Inserting into the
field equation (\ref{waveeqn},\ref{BOXD1}) yields the dispersion relation
\beq
0=-1+\frac1{d}\sum_i e^{-ik_i} + \frac1{d}\sum_i e^{ik_i}
-\frac1{d(d-1)}\sum_{i\ne j} e^{i(k_i-k_j)}.
\label{dr}
\eeq
For small $k_i$ (compared to $1/\epsilon$)
this reduces to $\sum_{i\ne j}k_i k_j=0$, which
says that $k_i$ is a null vector.
That is,
the continuum dispersion relation is recovered
in this limit.
The special cases where all but one of the $k_i$ vanish
are exact solutions both on the lattice
and in the continuum. These are plane waves
whose constant phase surfaces are level sets
of one of the $d$ null coordinates.

If the mode is written as
$\exp(-i\o t + i\vec{k}\cdot\vec{x})$, the
frequency $\o$ and spatial wave vector $\vec{k}$
are related to $k_i$ via
\beq
k_i=\o - (d-1)\vec{k}\cdot \hat{s}_i,
\eeq
where the unit vectors $\hat{s}_i$ point to the
corners of the spatial tetrahedron (see discussion
in section \ref{hd}).
We do not assume $\o$ is real since we wish
to discover whether or not there are
modes that grow or decay exponentially with time.
However, as in the continuum, we restrict to
real spatial wavevectors.

For general $k_i$ the dispersion relation (\ref{dr})
can be expressed as
\beq
\S(k)+\S(-k) =\frac{d\S(k)\S(-k) +d-2}{d-1},
\label{Sigmadr}
\eeq
where
\beq
\S(k)=\frac1{d}\sum_i e^{ik_i}.
\eeq
In terms of frequency and spatial wavevector
$\S$ can be factored as
\beq
\S(\o,\vec{k})=e^{i\o}z(\vec{k}),
\eeq
where
\beq
z(\vec{k})=\frac{1}{d}\sum_i e^{-i(d-1)\vec{k}\cdot \hat{s}_i}.
\eeq
Using the polar form $z(\vec{k})=r e^{i\theta}$,
the dispersion relation (\ref{Sigmadr}) becomes
\beq
\cos(\o+\theta) =\frac{dr^2 +d-2}{2(d-1)r}.
\label{polardr}
\eeq
Since the spatial vector is assumed real, $0\le r\le 1$.

In the two-dimensional case
({\ref{polardr}) reduces to $\cos\o=\cos k$,
where $k$ now refers to the spatial wavevector.
If $\cos k\ne0$ the general solution is
$\o=\pm k$, as in the continuum, corresponding to right and
left moving modes.
Equivalently, $k_2=0$ or $k_1=0$.
If $\cos k=0$, then the division
by $r$ in going from (\ref{Sigmadr}) to (\ref{polardr})
is disallowed,  and in fact $k=\pm\pi/2$ is also
a solution for any value of $\o$. These are spatially
alternating modes, since the spatial step length is $2\e$.
They are the pure gauge solutions discussed in Sec.~\ref{ivp},
which are excluded on a periodic lattice with an odd number
of points.

If the dimension $d$ is greater than two, then the right
hand side of (\ref{polardr}) is a positive function that
is concave upward and diverges as $r$ goes to zero or infinity.
This function is less than
unity only for $(d-2)/d<r<1$.  When the wavevector is small,
compared to the inverse lattice spacing,
$r$ is close to unity, so the frequency is real.
If it is possible to choose
$\vec k$ so that $r<(d-2)/d$, then there will exist a complex
solution to (\ref{polardr}), implying the
existence of exponentially growing or decaying modes
with wavelengths of order the lattice spacing.
We have not tried to determine whether there exist boundary
conditions that exclude such modes, but we now show by an example
that they can exist.

To simplify the computations
we choose the spatial wavevector to be
orthogonal to $\hat{s}_i$ for $i\ge 3$, so it
has the form $\a(\hat{s}_1-\hat{s}_2)$ for some
constant $\a$. We then find
$z(\vec{k})=[(d-2)/d] + (2/d)\cos\a d$. Hence
$r=|z|<(d-2)/d$  if $\cos\a d$ is negative.
For such values of $\a$, the mode is unstable.

On a compact lattice, such a mode must satisfy
periodic boundary conditions.
Consider a toroidal compactification of the lattice
as described at the end of Sec.~\ref{hd}, with identification
after $N$ lattice translations in each of the $(d-1)$
directions $\hat{s}_i-\hat{s}_1$ for $i=2,\dots d$.
The lattice step with $i=2$ for example is given by
$(v^1, v^2, v^3,\dots,v^d)\rightarrow(v^1-1, v^2+1, v^3,\dots,v^d)$.
In order for the mode function
to be periodic after $N$ steps in each of these
directions, its wavevector must satisfy
$k_i - k_1 = (d-1)\vec{k}\cdot(\hat{s}_1-\hat{s}_i)=2\pi n_i/N$.
With the above form of $\vec{k}$, these periodicity
conditions reduce to $2\a d=2\pi n_2/N$, and
$\a d=2\pi n_i/N$ for $i>2$. Thus we must choose
the $n_i$ equal to a common value $n$ for $i>2$,
and $n_2=2n$. This mode then satisfies the periodicity
condition if $\a d=2\pi n/N$. If we choose
$n$ so that $\cos 2\pi n/N$ is negative, this
mode is unstable.

We are not interested in trying to make sense of
the quantum theory in the presence of unstable modes.
Perhaps under appropriate boundary conditions they
are excluded. Certainly they can be eliminated
by adopting a different sort of discretization scheme.
The problem need not concern us here, since the
example we shall study in detail is two-dimensional.
Moreover, our general approach to quantizing the
field in the presence of mode birth is independent
of the issue of stability of the chosen finite difference
scheme.

In the continuum, for a fixed spatial wavevector there
are two frequencies that solve the
dispersion relation, $\o=\pm|\vec{k}|$.
On the hyperdiamond lattice, there are still two
frequencies (modulo $2\pi$) that solve the dispersion
relation, but, except in two dimensions, the
frequencies are not negatives of each other.
Indeed, if $(\o,\vec{k})$
is a solution to (\ref{polardr}),
then so is $(\o',\vec{k})$ if
\beq
\o'=-\o-2\theta.
\eeq
This is equivalent to the condition
$\S(\o',\vec{k})=\S(-\o,-\vec{k})=\bar\S(\o,\vec{k})$,
where the last equality holds only if $\o$ is real.

Another difference with the continuum in more than
two dimensions is that
if $(\o,\vec{k})$ is a solution, then $(\o,-\vec{k})$
is generally {\it not} a solution. Under
$\vec{k}\rightarrow-\vec{k}$ we have
$z(\vec{k})\rightarrow\bar{z}(\vec{k})$, so
$(r,\theta)\rightarrow(r,-\theta)$. Therefore
$(\o,-\vec{k})$ is a solution if and only if
$(\o-\theta)=\pm(\o+\theta)$, i.e. if $\theta=0$
or if $\o=0$.

As in the continuum, a massless scalar field has
a pair of spatially constant solutions on the
lattice.
To complete the list of modes we must include
these {\it zero-modes}.
The constant field with $k_i=0$ is a zero-mode,
as is the mode that grows linearly
in time, $\phi\propto\sum_i v^i$. These
play a role in the particle creation calculation carried
out below for the two-dimensional growing lattice.

\subsection{Quantization on a hyperdiamond lattice}

To carry out the Fock space quantization procedure
we must identify the decomposition of the space
of complex classical solutions into
positive and negative norm subspaces.
This yields the representation of the
algebra, and a pure state is specified using
an element of the Fock space.

\subsubsection{Harmonic modes}
The Klein--Gordon norm (\ref{KGip})
of the modes (\ref{hdmodes})
is
\beq
\la\eta_k,\eta_k\ra=2N |{\cal N}_k|^2{\rm Im}\S,
\eeq
where $N$ is the number of spatial points on
the lattice. We assume henceforth that all components of
the wavevector $k_i$ are real, so that
$\bar{\eta}_k\eta_k = |{\cal N}_k|^2$.
Modes with ${\rm Im}\S>0$ have positive
norm, and those with ${\rm Im}\S<0$ have negative
norm. The normalization factor
(up to an arbitrary phase) is
\beq
{\cal N}_k= \bigr(2N|{\rm Im}\S|\bigl)^{-1/2}.
\eeq
The space of modes thus decomposes into
these two subspaces, which are conjugate to each
other. Moreover, they are mutually orthogonal.
Consider the pairwise KG inner products:
\beq
\la\eta_k,\eta_l\ra=i\bigl(\bar\S(l)-\S(k)\bigr)
\sum_{p\in S}\bar\eta_k\eta_l.
\eeq
The sum will vanish unless the spatial wave vectors
for $k_i$ and $l_i$ are equal.
As seen in Sec.~\ref{sec:hdmodes},
for a given spatial wavevector the allowed frequencies
are $\o$ and $-\o-2\theta$, corresponding to
$\S(k)$ and $\S(k')=\bar\S(k)$. Thus
if the modes with the same spatial wavevector
are different, then $\S(k)=\bar\S(l)$. In this
case, one mode has positive norm and the other has
negative norm, and the inner product vanishes.
The modes thus provide a suitable orthonormal
basis of complex solutions
for constructing a Fock space representation
of the field algebra. We use the notation
$\xi_k$ for the positive-norm modes,
which are just those $\eta_k$ for which ${\rm Im}\S(k_i)>0$.
Explicitly, they are given by
\beq
\xi_k=\frac{1}{\sqrt{2N{\rm Im}\S(k_i)}}e^{-ik_i v^i}.
\label{nmodes}
\eeq

\subsubsection{Zero-modes}

The constant solution $\r=1$ and
the linear-in-time solution
$\s=t=\sum_i v^i$ both have vanishing norm,
and they are orthogonal to all the harmonic modes
$\eta_k$ for $k\ne0$.
The KG inner product between them
is $\la\r,\s\ra=iN$. This two-dimensional
zero-mode subspace can be split into a
positive-norm direction and its conjugate,
orthogonal, negative-norm direction, but there is
no unique way to do so. The different choices
of positive-norm zero-modes are $\r -i\g\s$,
where $\g=\g_R+i\g_I$
is a complex number with positive real part.
Explicitly, the normalized,
positive-norm zero-mode choices are
\beq
\xi_0^{(\g)}=\frac{1}{\sqrt{2N\g_R}}\Bigl(1-i\g\sum_i v^i\Bigr).
\label{zmode}
\eeq

\subsubsection{Field operator and Fock vacuum}
\label{sec:opvac}
The field operator is given as in (\ref{fieldop})
by
\beq
\hat\phi = \sum_k \bigl(\xi_k a_k + \bar\xi_k a^\dagger_k\bigr),
\label{fieldop2}
\eeq
where the sum is over those $k_i$ that satisfy the dispersion
relation with ${\rm Im}\S(k_i)>0$,
and the $a_k$ and $a_k^\dagger$ satisfy the
usual commutation relations of lowering and raising operators.

The Fock vacuum is the state satisfying $a_k|0\ra$
for all $k$. In this state, the expectation value of
the field operator vanishes, and that of the square
yields
\beq
\la0|\phi(p)\phi(p)|0\ra=(2N\g_R)^{-1}
\Bigl((1+\g_It)^2+(\g_R t)^2\Bigr)
+ \sum_{k\ne0}\Bigl(2N{\rm Im}\S(k)\Bigr)^{-1}.
\label{spread}
\eeq
The time dependence coming from the zero-mode shows that the
state is not time independent
except for $\g=0$, which does not correspond to a normalizable
zero-mode. The field amplitude of the zero-mode is like a
free particle, whose wavefunction always spreads.

One might be inclined to identify the state with the
lowest energy as the preferred vacuum. As for a free particle,
there is no normalizable state that corresponds to the lowest
energy of the zero-mode degree of freedom. Considering
just the other modes, there is also a fundamental issue
on the lattice: since time evolution is discrete, there
is no uniquely defined Hamiltonian, but rather a finite-time
evolution operator $U(\D t)$. Taking the logarithm one could
define a Hamiltonian $H=(i/\D t)\ln U$, but this is ambiguous due
to the freedom to add a multiple of $2\pi/\D t$ to any eigenvalue
without changing the evolution operator.

Even if there is no meaningful lowest energy state, one can
still ask if the Fock vacuum is an eigenstate of the
evolution operator. Other than for the zero-mode,
the answer is yes: if the evolution operator $U$ is defined by
$U^\dagger\hat{\phi}(v^1,v^2,\dots,v^d)U
=\hat{\phi}(v^1+1,v^2,\dots,v^d)$
then, omitting the zero-mode, $U|0\ra=|0\ra$ (up to an arbitrary phase).
Combining such translations of the null coordinates one
obtains all space and time translations.

\section{Mode birth on a growing two-dimensional lattice}
\label{example}

To explore the phenomenon of mode birth in a simple
setting we now specialize to two-dimensional lattices.
We consider a model in which the lattice is initially
regular and then some points are added, after which
the lattice is again regular. We first specialize
the results of the previous section to two-dimensional
regular lattices, then turn to the case where points are added.

\subsection{Regular diamond lattice}

On a regular diamond lattice in two dimensions
the field equation (\ref{waveeqn},\ref{BOXD1}) can be written as
\beq
\Bigl[\bigl(\phi(\up)+\phi(\dop)\bigr) - \bigl(\phi(\lp)+\phi(p)\bigr)\Bigr]
+
\Bigl[\bigl(\phi(\pu)+\phi(\pd)\bigr) - \bigl(\phi(p)+\phi(\pr)\bigr)\Bigr]
=0,
\label{2deq}
\eeq
where $\up$ and $\pu$ are the left and right children,
$\dop$ and $\pd$ are the left and right parents,
and $\lp$ and $\pr$ are the
left and right siblings of $p$.
In terms of the
{\it diamond operator},
\beq
\Diamond\phi(p)=\bigl(\phi(\up)+\phi(\dop)\bigr) -
\bigl(\phi(\lp)+\phi(p)\bigr)\Bigr),
\eeq
the field equation (\ref{2deq})
takes the form
\beq
\Diamond\phi(p)+\Diamond\phi(\pr)
=0.
\label{2deqD}
\eeq
This asserts that
the sum of every pair of adjacent diamonds
vanishes, so the diamonds on each time slice
have the same magnitude and alternating sign.
On a finite periodic lattice with $N$ points
there are $N$ diamonds. If $N$ is odd the
alternation is consistent only if each diamond
vanishes individually.

The mode functions were found in Sec.~\ref{sec:hdmodes}.
For the right-moving modes ($k_2=0$) we have
$2{\rm Im}\S=\sin k_1$. The positive-norm
modes (\ref{nmodes}) are those for which
$\sin k_1>0$, and are given explicitly by
\beq
\xi_k=\frac{1}{\sqrt{N\sin k_1}}e^{-ik_1v^1}.
\label{2dnmodes}
\eeq
A similar expression with $(k_1,v^1)$ replaced by $(k_2,v^2)$
yields the left-movers.
The zero-mode (\ref{zmode}) takes the same form in
all dimensions.

As explained in section \ref{ivp}, we restrict to
odd values of $N$ to avoid the alternating gauge
freedom. Periodicity for the mode solutions
(\ref{2dnmodes}) implies that $k_1=2\pi n/N$ for some
integer $n$,
so the spectrum of modes is discrete. We obtain the
complete set if $n$ runs from $-(N-1)/2$ to $(N-1)/2$.
The positive-norm modes are those with positive $k_1$
in this range, and the time-independent
zero-mode corresponds to $n=0$.
There are $N$ positive-norm modes all together:
$(N-1)/2$ left-movers, $(N-1)/2$ right-movers,
and a zero-mode.

\subsection{Growing lattice: mode birth}

We now consider a lattice which is irregular
in a localized region where some points
are added. To preserve an odd number of points
an even number must be added on each slice.
We add just two points, with the connectivity\footnote{We
also investigated the case where
$e$ and $f$ are children of $h$, which
yielded similar results. It might
be interesting to explore the consequences of more
complicated connectivity of the birth points.}
illustrated in Fig.~\ref{birth}.
\EPSFIGURE[pos]{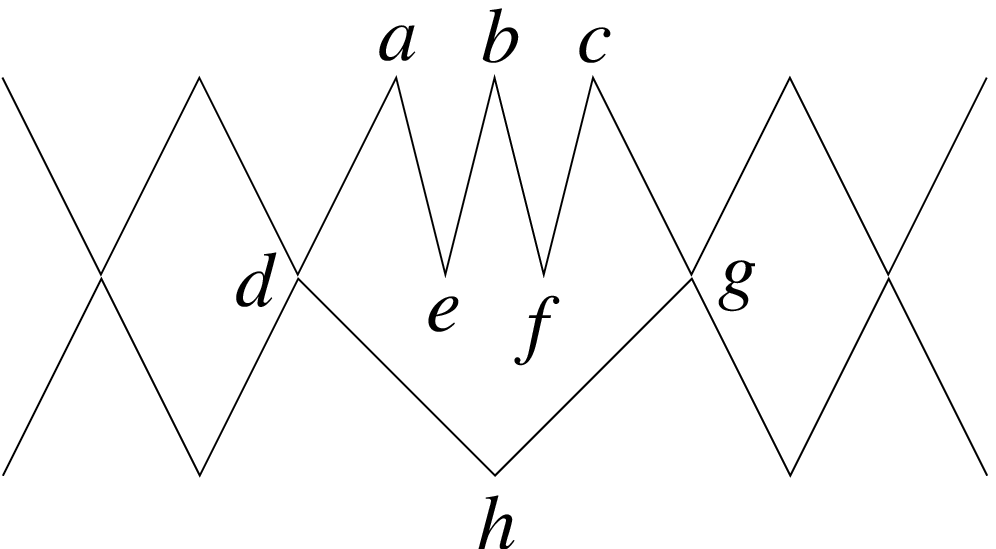, width=6cm}{ \label{birth}
\small Structure of the lattice around the birth
event.
\smallskip}
The point labels $a,b,c,d,e,f,g,h$ will be
referred to in this section.
The equation
of motion from varying the field at the \textit{birth
point} $e$ implies that
\beq
\phi(e)=\bigl(\phi(a)+\phi(b)\bigr)/2,
\label{birtheq}
\eeq
i.e. $\phi(e)$ is equal to the average of the
future linked values. A similar
equation holds for the birth point $f$.

Evolution of solutions backward through the birth event,
if possible, is unique, as required by our linear
phase space construction
described in section~\ref{Gamma}.
Evolution forward is always possible, but is
not unique since the values at the newborn
points are undetermined. All such evolutions
correspond to points in the phase space.

We call the slice containing the birth points the
\textit{birth time}.
 Evolving backwards,
the values at the birth points can be determined from
those on the subsequent two slices, but unless it
so happens that the birth point
values are the average of their
children's values, the field equation at the
birth points is not satisfied. Put differently,
there may be no way to take the
next step backwards that satisfies the field
equation at the birth time. Hence,
there are half-solutions
(cf.~section~\ref{Gamma}).  The initial data for
these \textit{birth solutions}
are given at the birth time and on the subsequent slice.

The general definition of the
quantum theory in
section~\ref{Q} applies to this growing lattice.
The phase space includes the whole solutions
together with the half-solutions.
The algebra of linear observables on this phase
space is canonically quantized, yielding the quantum
field algebra. Together with a choice of state---i.e.
a positive linear functional on this algebra---the
definition of the quantum theory is
complete. To construct a concrete representation,
define a particular state, and compute the
value of some observables, we now use the method of
mode decompositions.

\subsection{Mode decompositions and Fock space}

After the birth event the lattice is
regular, and therefore the Fock space
can be based on the harmonic modes (\ref{2dnmodes})
together with the zero-mode (\ref{zmode}).
A simple set of observables to examine
are the number operators $N_k=a_k^\dagger a_k$
for these `out' modes. We shall compute the
expectation values $\la 0|N_k|0\ra$,
taking for the state $|0\ra$
the Fock vacuum based on a set of `in' modes.

To find a complete set of in-modes we begin
with the modes before the birth event, and choose a
particular evolution among the many that carry
these modes through the birth event.
We then determine the half-solutions that
are orthogonal to the evolved in-modes,
and augment the Fock space with these.
This requires only the values of all modes
at the birth time and the subsequent slice.

It is worth emphasizing that
the arbitrary choice resolving the
ambiguity of evolution through the
birth event does not affect the quantum
algebra of observables, since that is constructed
from {\it all} solutions and half-solutions.
Nevertheless, the evolution rule does affect the inner
product of the evolved pre-birth-modes with
the half-solutions, since the inner products
must be evaluated on or after the birth slice.
One cannot evaluate on a slice prior to the
application of the arbitrary rule, since
the birth solutions are undefined there.

Although it does not affect the algebra
as a whole, the evolution rule
does have a physical impact
in our construction through the definition of
the Fock vacuum. The Fock space is built
from an orthonormal basis of positive-norm modes
that are orthogonal to their conjugates.
The evolution rule affects the construction
of this basis, so it affects the Fock vacuum.
Since all representations are unitarily
equivalent however, any Fock state defined using one
evolution rule can be expressed as a Fock state
in a different Fock space constructed with
a different evolution rule. Thus we are not
choosing between {\it theories}, but rather
just between equivalent {\it representations}.

\subsubsection{Evolution rule}
\label{rule}

For simplicity here we choose the evolution
rule that sets the evolved field to zero at the
birth points $e$ and $f$. It might seem more natural to linearly
interpolate the field between the neighbors
$d$ and $g$ of the
birth points. However, as just explained, this would change
only the Fock vacuum, not the theory itself.
We did examine the particle production using the
Fock vacuum arising from the ``interpolation rule",
and the results were similar to those of the
``zero rule". Since the details are simpler
for the zero rule, we adopt it for expository
purposes.

Let us now see how the fields evolve through
the birth event using this evolution rule.
With $\phi(e)=0$,
the field equation (\ref{birtheq})
at the birth point $e$
implies that $\phi(a)=-\phi(b)$.
Similarly the equation at $f$ implies
that $\phi(c)=-\phi(b)$. Taken together
these imply that $\phi(a)=\phi(c)$.

Now consider the rest of the equations at
the birth time.
The field equation at the neighbors
$d$ and $g$ differs from
the standard form (\ref{2deqD})
only in the shape of one of the
two diamonds.
In particular, in
the equation from the variation of the
value $\phi(d)$,
the diamond on the right consists of
the vertices $(d,a,g,h)$, while
for the variation of $\phi(g)$
the diamond on the left consists
of $(d,c,g,h)$. These two
diamonds agree except at the
top vertices $a$ and $c$.
The diamonds alternate sign around the
compact lattice, and there are
an even number $N+1$ of diamonds (since in
effect the middle diamond has split into
two, hence the two middle diamonds
must be opposite). But we already
saw that the field equations
at the birth points imply
$\phi(a)=\phi(c)$, so these diamonds
must agree. The only consistent
solution is that all the diamonds
vanish. With this information
we can now easily evolve the pre-birth
harmonic modes and zero-modes to the next
slice after the birth time.

\subsubsection{Choice of birth modes}
\label{choice}

There are two birth points, so there are four
independent real birth modes. We wish to choose
these so they are orthogonal to the evolved
pre-birth modes. The birth modes are to
be characterized by their initial data
at the birth time, i.e. their values at the birth
time and on the next slice.

To identify these birth modes, we
first note that any mode
whose initial data is non-vanishing
only at the birth points $e$ and $f$
is orthogonal to any evolved pre-birth
mode. This is because (by the evolution rule)
the latter vanish at $e$ and $f$, and also,
by the field equation
(\ref{birtheq}), the averages on $IF(e)$ and
$IF(f)$ vanish.
For two of the birth modes we thus take
the symmetric mode $S_1(e)=S_1(f)=1$
and the anti-symmetric
mode $A_1(e)=-A_1(f)=-1$, where the initial data
vanishes at all other points.

For the other two birth modes we set the
data to zero at the birth time.
The symmetric mode with
$S_2(b)=1$ (and zero elsewhere)
is orthogonal to all pre-birth-modes
since the point $b$ is not to the future
of any point except $e$ and $f$.
The fourth and final birth mode
is the anti-symmetric mode defined
by $A_2(a)=-A_2(c)=-1$, $A_2(b)=0$, and
the requirement that
$A_2$ alternates between $+1$ and $-1$ on
the other points around the slice.
This is orthogonal to any pre-birth-mode since
the sum of the values to the future
of any birth time point other than
the $e$ and $f$ vanishes.

Being real, the four birth modes $S_{1,2}$
and $A_{1,2}$ have vanishing norm,
and the non-zero inner products
between them are $(S_1,S_2)=(A_1,A_2)=i$.
We seek two complex linear combinations that
have positive norm and are
orthogonal to both of their conjugates.
The two modes
\beq
S=(S_1 -iS_2)/\sqrt{2},
\qquad A=(A_1 -iA_2)/\sqrt{2},
\label{bmodes}
\eeq
have positive-norm, and
satisfy the orthogonality
requirement, since
all modes are orthogonal to their own
conjugate, and symmetric modes are orthogonal
to antisymmetric ones.

Other choices than (\ref{bmodes})
could be made. Like for the
zero-mode (\ref{zmode}), the factor of
$i$ could be replaced by
$i\g$ (and the denominator
replaced by $\sqrt{2{\rm Re}\g}$).
Moreover, since there are two
independent positive-norm modes,
there is further freedom that involves
mixing between them. The different
Fock vacua so obtained are
called two-mode squeezed states.
For the present model we shall
make the simple choice
given by (\ref{bmodes}).

\subsection{Growth-induced particle creation}
Now we have all the ingredients for the particle
production computation. We wish to compute the
mean occupation number of each out-mode when
the state is the Fock in-vacuum.

The out-number operator is
$N_{k,{\rm out}}=a_{k,{\rm out}}^\dagger a_{k,{\rm out}}$.
To compute its mean value one needs to express
the annihilation operator $a_{k,{\rm out}}$
for the out-mode $\chi_k$ in terms of the in-modes $\xi_i$.
We have
\bea
a_{k,{\rm out}}&=&\la\chi_k,\hat{\phi}\ra\\
&=&\sum_i \Bigl(\la\chi_k,\xi_i\ra a_i +
\la\chi_k,\bar\xi_i\ra a_i^\dagger\Bigr),
\eea
and using the Fock vacuum condition $a_i|0_{\rm in}\ra=0$ and the
commutation relations (\ref{aadagger}) this yields
\beq
\la 0_{\rm in}|N_{k,{\rm out}}|0_{\rm in}\ra
=\sum_i |\la\chi_k,\bar\xi_i\ra|^2.
\label{numba}
\eeq
The sum is over the ingoing harmonic modes and zero-mode, as well
as the birth modes. Formulae for the Bogoliubov coefficients
$\b_{ki}= \la\chi_k,\bar\xi_i\ra$ are given in the appendix.

\EPSFIGURE[pos]{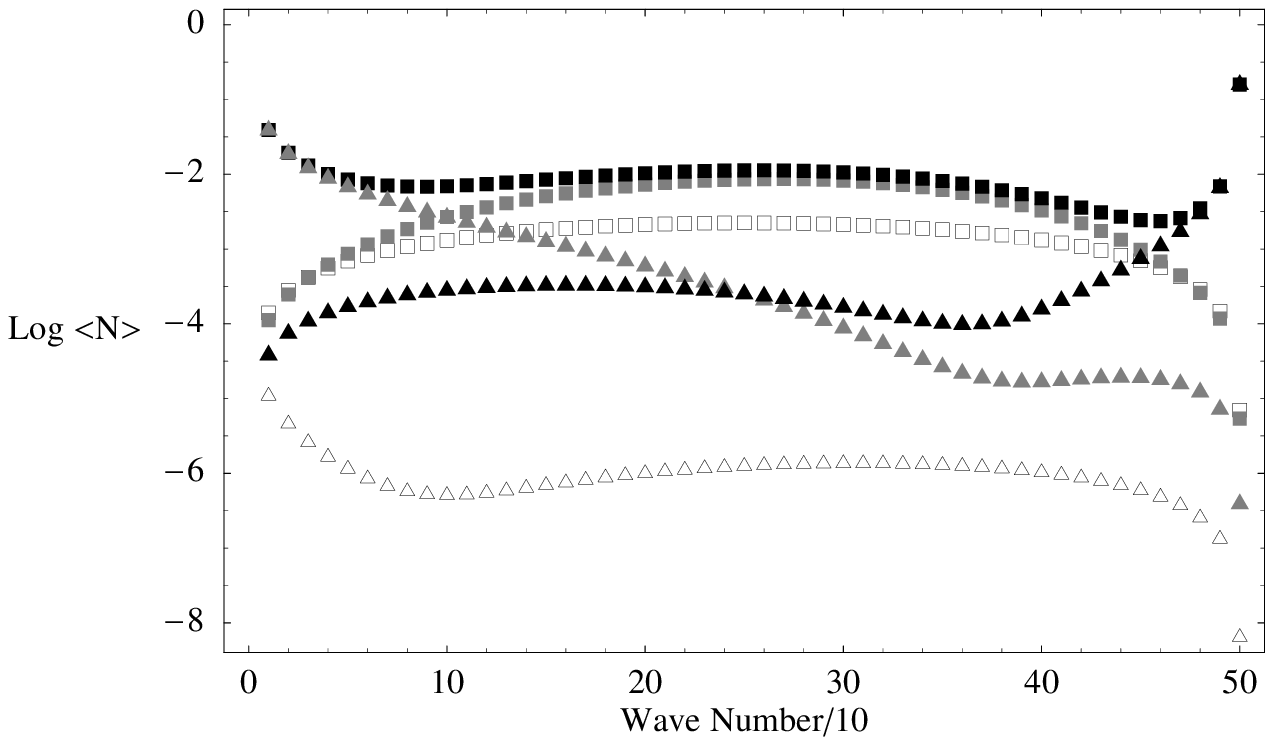, width=8cm}{\label{spectrum}
\small Spectrum of created particles on a lattice that
begins with 999 points and ends with 1001 points
after a birth process as depicted in Fig.~\ref{birth}.
Plotted is the
Log of the mean number of left-moving particles vs. $n/10$,
where the wave number is $k=2\pi n/1001$.
Black squares give the total, while the other curves show the
contributions from the left- and right-movers  (grey and unfilled squares),
the symmetric and anti-symmetric birth modes (grey and black triangles),
and the zero-mode (unfilled triangles). The zero-mode parameter is
$\g=0.4$.
\smallskip}
The spectrum of left-moving created particles is displayed
in Fig.~\ref{spectrum},
along with the contributions to the mean
number (\ref{numba}) from the left-movers, right-movers,
zero-mode, symmetric birth mode, and anti-symmetric birth mode.
The number of lattice points starts out as $999$, and
the zero-mode parameter $\g=0.4$ in (\ref{zmode}) has been chosen.
Most of the contribution comes from the symmetric birth mode
at low wavenumbers, from the anti-symmetric birth mode
at high wave numbers, and  from the harmonic modes
at intermediate wave numbers. The contribution from left-moving
harmonic modes is several times higher than from right-moving ones
except at very low and very high wave numbers.
The contribution from the zero-mode is
far smaller than all the rest.

The number itself is presumably not as relevant as the
{\it energy} of the created particles. As discussed
in section~\ref{sec:opvac}, the Hamiltonian---and
therefore the energy---is not well-defined on a lattice,
due to the ambiguity of adding any multiple of
$2\pi/\epsilon$ to eigenvalues of the Hamiltonian
without altering the evolution operator.  Nevertheless,
we presume that, as a model for the microstructure of
spacetime, the degree of excitation relative to the
translation-invariant Fock vacuum is relevant in
determining the gravitational back-reaction.
We further presume that
the smallest positive value of the
frequency for a given positive-norm
mode would determine its gravitating energy,
at least for modes with
frequency much less than $2\pi/\epsilon$ (mod $2\pi/\epsilon$).

This
energy and its $N$ dependence can be understood
roughly from the
form of the Bogoliubov coefficients in the
appendix as follows.
The squared Bogoliubov coefficients (\ref{LL},\ref{RL})
for harmonic in- and out-modes (\ref{nmodes})
goes as $|\b_{kl}|^2\sim N^{-2}(\sin k/ \sin l)$.
The sum over $l$ values gives $\sum_l (1/\sin l)= O(N\ln N)$,
so the contribution to the number from harmonic modes goes
as $(\ln N/N)\sin k$. A given $k$ mode carries an energy $k$.
(Possibly we should instead use $\sin k$, but that difference
will not qualitatively affect the result.) Multiplying the number by
$k$ and summing over $k$ values yields $\sum k \sin k =O(N)$,
hence the total energy contribution from the harmonic in-mode
terms is $O(\ln N)$.

For the symmetric birth mode the
Bogoliubov coefficient goes as $|\b_{kS}|^2\sim (N\sin k)^{-1}$
as long as $k$ is not too close to $\pi$.
Multiplying by the energy $k$ then yields the sum
$\sum_k (k/\sin k) = O(N)$, hence the contribution to the
energy is $O(1)$. A similar result holds for
the antisymmetric birth mode, with a suppression from $k$ near zero.
The squared zero-mode Bogoliubov coefficient is suppressed
by an additional factor of $1/N$, yielding an energy contribution
of order $O(\ln N/N)$.

The total created energy due to the birth event is thus of
order $\e^{-1}\ln N $, where the dimensionful constant
$\e$ setting the scale of length is restored. If
every unit of spatial length is born accompanied by
this amount of energy creation, the result would be incompatible
with the effective field theory notion that the field remains
near the adiabatic ground state. A doubling of the size of
the universe would produce a Planck energy density of particles
in this one-dimensional model.

It is possible as usual to express the out-vacuum
$|0_{\rm out}\ra$ as a squeezed state
in the Fock space of the in-modes. If $\a_{ki}$ and
$\b_{ki}$ are the Bogoliubov coefficients defined by
$a_{k,{\rm out}}=\sum_i (\a_{ki} a_i + \b_{ki} a_i^\dagger)$,
this state takes the form
$\exp(-\a^{-1}_{ik}\b_{kj}a_i^\dagger a_j^\dagger)|0_{\rm in}\ra$,
where repeated indices are summed.
In this squeezed state the in-modes are excited and
there are specific correlations between the excitations
of the different in-modes.

If the ingoing harmonic
modes are assumed to be in their ground state, however, no
amount of fiddling with the birth mode states and the correlations
between those and the harmonic modes can substantially change
the amount of energy created. The out-annihilation operator
can be expanded as
\beq
a_{k,{\rm out}}=\a_{kl} a_{l,{\rm in}} + \b_{kl} a^\dagger_{l,{\rm in}}
+ \a_{kB}a_B + \b_{kB} a^\dagger_B,
\eeq
where both birth modes are lumped into the one index $B$.
Assuming only that the in-state is the vacuum for
the harmonic (and zero-) modes, $a_{l,{\rm in}}|0\ra=0$,
the mean number evaluates to
\beq
\la 0_{\rm in}|N_{k,{\rm out}}|0_{\rm in}\ra
=\sum_l |\b_{kl}|^2+ \|( \a_{kB}a_B + \b_{kB} a^\dagger_B)|0\ra\|^2.
\label{numba2}
\eeq
The birth mode state could be chosen so that the second term in (\ref{numba2})
vanishes, but this would not have a substantial impact on the order of
magnitude of the injected energy. The spectrum in Fig.~\ref{spectrum}
shows that, by themselves, the harmonic modes already contribute an
order unity fraction.

Without modifying the assumption that there are no incoming
particles,
the only way to change this conclusion is to modify the
evolution rule. This rule affects the in-Fock representation,
so it affects the state corresponding to the in-Fock vacuum.
The expectation value of the out-number-operator (\ref{numba2})
is sensitive to this via the Bogoliubov coefficients
$\b_{ki}= \la\chi_k,\bar\xi_l\ra$, since the out-modes $\chi_k$
are not defined until the birth time. Some of these coefficients
can be suppressed by adjusting
the evolution rule.
One could tune the rule to kill individual coefficients,
or one could choose a ``smoothing"  interpolation rule
that would suppress the contributions from low wave numbers.
However, it seems clear that there is no way to avoid
the $O(\e^{-1})$ energy injection.

\section{Discussion}

Using the notion of a phase space of half-solutions,
we have managed to quantize a linear field theory on
the background of a lattice with a growing number of points.
To ensure a well defined symplectic structure on a linear
phase space the lattice
was assumed to be layered by time slices, and to
admit at most one backwards evolution, and at least
one forward evolution from any initial data. We were forced
to this rather special structure by the requirement of
staying very close to standard linear QFT.
It may be possible to generalize the structure in
various ways and still make sense of the quantum theory.
One obvious small generalization would be to a different
type of lattice that is still layered.

The computation of particle creation in the
two-dimensional model showed that the addition
of two points necessarily injects an energy of order
$\e^{-1}$ into the field. If more points are added
at different times,
a similar energy is injected for every unit
of length $\e$ added. Since the in-state is no
longer the in-vacuum, the particle production would
not be identical, but it must clearly be
of the same order of magnitude.
This would produce an enormous energy
density of order $\e^{-2}$ during expansion of a
model two-dimensional universe.
Extrapolating to our four-dimensional
universe, this behavior is not observed.
One might hope that in four dimensions the
effect of adding points at the lattice scale would not
be so brutal, but this seems highly unlikely,
as the form of the Bogoliubov coefficients  should
be comparable to those in the
appendix.\footnote{Perhaps the extra $\ln N$ would
go away in higher dimensions, but that is not our first
concern.} This should be expected, since
the birth process violates time and space translation
symmetry on a time and length scale of $\e$, hence it
it should be accompanied by a violation of
energy conservation of order
$\e^{-1}$.

A way out of this energy catastrophe
might arise if many points are added
at the same time. Then many modes are born at the
same time, and wave interference could perhaps
suppress the net energy injected. This question
can be studied in the two-dimensional model.

If interference effects cannot reduce the injected
energy, then one would conclude
that QFT on a {\it background} growing lattice
is not the right setting to describe mode birth
in a growing universe.
If there is nevertheless a physically correct formulation,
it would presumably operate
at the level where the spacetime itself is treated as dynamical,
rather than as a fixed background. The complete dynamics
might ensure that the system remains
near the adiabatic vacuum.
Perhaps this scenario could be explored using
the causal dynamical triangulations formulation
of quantum gravity\cite{Loll}.

Apart from trying to address the fundamental problem
of the cosmological
vacuum, our formulation
of mode birth could be applied
to field propagation outside a black hole.
In a geodesic normal coordinate system the spatial
metric grows with respect to the geodesic proper time,
by an amount that depends upon the radius.
Discretizing the radial coordinate and keeping time
continuous one obtains an expanding lattice
version of the black hole spacetime
in which the number of degrees of
freedom is constant. The Hawking effect on such
a falling spatial lattice was studied in
\cite{Corley:1997ef,Jacobson:1999ay}.
It was found that the outgoing
modes arise via a process analogous to Bloch oscillation,
and the continuum Hawking effect is recovered when the lattice
spacing is small compared to the inverse surface gravity of
the horizon. Rather than keeping time continuous
and allowing the lattice to expand, one could
instead discretize the time and allow for points
to be added so as to keep the lattice scale roughly
constant. This would bring into play the sort of
indeterminate evolution and mode birth studied here.
In spite of the energy injection by
mode birth, the Hawking effect could
presumably be recovered provided the Hawking
occupation number per outgoing mode is greater than
the number induced by the birth events.

\acknowledgments

This work was supported in part by the NSF under grant
PHY-0300710 at the University of Maryland, and by the CNRS
at the Institut d'Astophysique de Paris.

\appendix

\section{Bogoliubov coefficients}

In this appendix we present the inner-products between harmonic out-modes
and the various negative-norm in-modes
that appear in the two-dimensional, growing lattice
example of Section~\ref{example}.

For our purposes---calculation
of the value of the number operator---we need only the norm
of the value of the inner products.  Consequently, we can
freely choose the over-all phase of a given mode.
For definiteness, here we
set the phase of the out-modes to zero at point $d$ (see
Figure~\ref{birth} for point labels).
For the harmonic, non-zero in-modes, we have set the phase
to zero at point $h$.  The left-moving mode, when
evolved according to the rule of Section~\ref{rule}, has zero
phase at  point $d$, the left child of $h$.  The right-moving
mode has zero phase at point $g$, the right child of $h$.
We take the
zero-mode and the birth modes exactly as defined in
Section~\ref{choice}.

The formulas correspond to a left-moving harmonic out-mode
$\chi_k$
(we do not consider the out-zero-mode).  The results for a
right-moving mode are identical up to a phase, except that
Eqn.~\eqref{LL} now corresponds to the inner product with a
right-moving in-mode, and Eqn.~\eqref{RL} to that with a
left-moving in-mode.  All formulas hold for either sign
of $k$; i.e.~for both positive- and negative-norm out-modes.
The positive-norm modes correspond to those for which
$\sin(k) > 0$, and vice-versa for the negative-norm modes.

For the inner-product with
a left-moving in-mode of wave-number $l$, $\xi_l^L$,
we have:
\be
\label{LL}
\begin{split}
    \la\xi^L_l,\chi_k\ra
                    &=A_{N-2}^l A_N^k\sin{k}
                        \Bigl(e^{i (l-k)}+
                        \sum^{N-3}_{n=0}e^{i n(l-k)}\Bigr)\\
        &=A_{N-2}^l A_N^k\sin k
        \Bigl(e^{i(l-k)} +
            \frac{1-e^{i(N-2)(l-k)}}{1-e^{i(l-k)}}\Bigr),
\end{split}
\end{equation}
with $A_N^k = 1/\sqrt{N |\sin{k}|}$.
  For a right-moving in-mode with
wave-number $l$, $\xi_l^R$, we have:
\be
\label{RL}
    \la\xi^R_l,\chi_k\ra
        = -e^{i(l-2k)}A_{N-2}^l A_N^k\sin{k}.
\ee
The above hold for both positive- and negative-norm in-modes, which
one
can classify by the sign of sine of the wave-number.

For the negative-norm in-zero-mode, $\xi_0^-$, with
free-parameter $\gamma$, we have:
\be
\label{ZL}
    \la\bar\xi_0^{(\g)},\chi_k\ra = e^{-i2k}
            \Bigl(\frac{A_N^k}{\sqrt{2\gamma(N-2)}}\Bigr)
            \bigl((-1+i\g)\sin k + 2\g e^{ik/2}\cos\tfrac{k}{2}\bigr).
\ee
For the negative-norm, symmetric birth mode $\bar S$, we have:
\be
\label{SL}
    \la\bar{S},\chi_k\ra = e^{-i2k}
            \Bigl(\frac{A_N^k}{\sqrt{2}}\Bigr)
        \bigl(2i\cos^2(k/2)-e^{ik/2}\cos(k/2)\bigr).
\ee
For the negative-norm, anti-symmetric birth mode $\bar{A}$, we have:
\be
\label{AL}
    \la\bar{A},\chi_k\ra  =e^{-i2k}
                    \Bigl(\frac{A_N^k}{\sqrt{2}}\Bigr)
                    \bigl(\sin k +ie^{ik/2}\sin(k/2)\bigr).
\ee
To obtain the inner-products with the positive-norm
birth or zero-modes, one can use the fact that,
from the definition of the inner-product,
$\la\phi,\psi\ra =  -\la\phi^* ,\psi^* \ra^*$,
where * denotes complex conjugation.
For example, $\la S,\chi_k\ra = -\la \bar{S},\chi_{-k}\ra^*$.

\end{document}